\documentclass[a4paper]{jpconf}
\usepackage{graphicx}

\def\ga{\mathrel{\mathchoice {\vcenter{\offinterlineskip\halign{\hfil
$\displaystyle##$\hfil\cr>\cr\sim\cr}}}
{\vcenter{\offinterlineskip\halign{\hfil$\textstyle##$\hfil\cr>\cr\sim\cr}}}
{\vcenter{\offinterlineskip\halign{\hfil$\scriptstyle##$\hfil\cr>\cr\sim\cr}}}
{\vcenter{\offinterlineskip\halign{\hfil$\scriptscriptstyle##$\hfil\cr>\cr
\sim\cr}}}}}

\begin{document}
\title{Ultra High-Energy Cosmic Ray Observations}

\author{Karl-Heinz Kampert}

\address{Bergische Universit\"at Wuppertal, Fachbereich C - 
Physik, Gau{\ss}str.\ 20, D-42119 Wuppertal}

\ead{kampert@uni-wuppertal.de}

\begin{abstract}
 The year 2007 has furnished us with outstanding results about the origin of the most energetic cosmic rays: a flux suppression as expected from the GZK-effect has been
 observed in the data of the HiRes and Auger experiments and
 correlations between the positions of nearby AGN and the arrival
 directions of trans-GZK events have been observed by the Pierre
 Auger Observatory.  The latter finding marks the beginning of
 ultra high-energy cosmic ray astronomy and is considered a major
 breakthrough starting to shed first light onto the sources of
 the most extreme particles in nature.  This report summarizes
 those observations and includes other major advances of the
 field, mostly presented at the 30$^{\rm th}$ International
 Cosmic Ray Conference held in M\'erida, Mexico, in July 2007.
 With increasing statistics becoming available from current and
 even terminated experiments, systematic differences amongst
 different experiments and techniques can be studied in detail
 which is hoped to improve our understanding of experimental
 techniques and their limitations.
\end{abstract}

\section{Introduction}
Understanding the origin of the highest energy cosmic rays is one
of the most pressing questions of astroparticle physics.  Cosmic
rays with energies exceeding $10^{20}$ eV have been observed for
more than 40 years (see e.g.\ \cite{Nagano-Watson}) but due to
their low flux only some ten events of such high energies could
be detected up to recently.  There are no generally accepted
source candidates known to be able to produce particles of such
extreme energies.  Moreover, there should be a steeping in the
energy spectrum near $10^{20}$ eV due to the interaction of
cosmic rays with the microwave background radiation (CMB).  This
Greisen-Zatsepin-Kuzmin (GZK) effect \cite{GZK} severely limits
the horizon from which particles in excess of $\sim
6\cdot10^{19}$ eV can be observed.  For example, the sources of
protons observed with $E\ge 10^{20}$ eV need to be within a
distance of less than 50 Mpc \cite{Harari-06}.  The
non-observation of the GZK-effect in the data of the AGASA
experiment \cite{Takeda-03} has motivated an enormous number of
theoretical and phenomenological models trying to explain the
absence of the GZK-effect and has stimulated the field as a
whole.  Only this year, with the final analysis of the HiRes-data
\cite{Abbasi-07a} and the advent of high-statistics and high
quality hybrid data from the Pierre Auger Observatory (PAO)
\cite{Yamamoto-07}, the situation has changed considerably: a
suppression such as expected from the GZK-effect is now observed
with high statistical significance.  The very recent breaking
news about the observation of directional correlations of the
most energetic Pierre Auger events with the positions of nearby
AGN \cite{Abraham-07} complements the observation of the GZK
effect very nicely and provides evidence for an astrophysical
origin of the most energetic cosmic rays.  Another key observable
allowing to discriminate different models about the origin of
high-energy cosmic rays is given by the mass composition of
cosmic rays. Unfortunately, such measurements are much more
difficult due to their strong dependence on hadronic interaction
models.  Only primary photons can be discriminated safely from
protons and nuclei and recent upper limits to their flux largely
rule out top-down models, originally invented to explain the
apparent absence of the GZK-effect in AGASA data.

In this article, prepared for the TAUP conference in Sendai
(Japan), we describe the status of each of these topics, as
reported during the recent International Cosmic Ray Conference
held in M\'erida, Mexico, in July 2007 (ICRC2007) and in
publications becoming available since.

\section{Experiments and their Exposures}

Most of the data available today at energies above $\sim
10^{18}$~eV are provided by the 100\,km$^{2}$ AGASA array
\cite{Takeda-03}, the HiRes fluorescence telescopes
\cite{Abbasi-07a} and the 3000 km$^{2}$ PAO \cite{Auger-NIM}.
Both, the AGASA and HiRes instruments have now closed while the
PAO started operation during its construction phase in 2004.
Even though its last surface detector stations are planned to be
deployed in March 2008 only, it has already accumulated the
largest exposure available today and it will continue to deliver
more than about 7000 km$^{2}$\,sr for each year of operation.
Table \ref{tab:expts} lists for various experiments the
approximate accumulated exposures and currently observed numbers
of events claimed to be above 10 and 50 EeV, respectively.  The
table (based on \cite{Watson-05}) includes for comparison smaller
air shower arrays, such as the phased out Haverah Park
\cite{Lawrence-91} and AKENO \cite{Nagano-92} arrays, the Yakutsk
\cite{Glushkov-05}, and the KASCADE-Grande array
\cite{Haungs-07}.  The Telescope Array (TA), led by a Japanese
consortium in Millard County, Utah, USA, has just begun operation
this year \cite{Fukushima-07}.  Like the PAO, the TA is a hybrid
detector.  It covers an area of 860 km$^{2}$ and comprises 576
scintillator stations and three FD sites on a triangle with about
35 km separation each equipped with 12 fluorescence telescopes.

The aperture assumed in calculating each exposure is appropriate
to about 50 EeV. Please note that the apertures of fluorescence
telescopes such as employed by HiRes are a growing function of
energy while those of ground arrays are flat above their
respective threshold energies.  A cursory comparison of the rate
of events above 50 EeV (the energy at which the exposures in
Table \ref{tab:expts} are calculated) makes clear that the
differences between the integral rates are much larger (by more
than a factor of 2) than can be accounted for by Poissonian
variations.  Possible reasons for the discrepancies will be
discussed below.

\begin{table}[h]
\label{tab:expts}
\caption{Exposure and approximate event numbers from various instruments.}
\footnotesize\rm
\begin{center}
\begin{tabular}{llrrcl} \br
{\bf Experiment} & {\bf status} & {\bf km}{\boldmath $^{2}$} 
     {\bf sr yr} & \multicolumn{2}{c}{\bf \# events} & {\bf based on}\\
 &  & @ {\bf 50 EeV} & {\boldmath $> 10$} {\bf EeV}
 & {\boldmath $> 50$} {\bf EeV} & {\bf Ref.}\\ \mr
Haverah Park  & 1962-1987    & $\sim 245 $ & 106 & 10 & \cite{Lawrence-91}\\
Yakutsk       & 1974-present & $\sim 900 $ & 171 &  6 & \cite{Watson-05,Glushkov-05}\\
AGASA         & 1993-2005    & 1620        & 886 & 46 & \cite{agasa-hp}\\
HiRes-I mono  & 1997-2006    & $\sim 4500$ & 561 & 31 & \cite{Bergman-07a,Bergman-07b,Stokes-07}\\
HiRes-II mono & 1999-2006    & $\sim 1500$ & 179 & 12 & \cite{Bergman-07a,Bergman-07b}\\
HiRes stereo  & 1999-2006    & $\sim 2400$ & 270 & 11 & \cite{Hanlon-07,Sokolsky-07}\\
Auger         & 2004-present & $\sim 7000 $& 1644 & 38 & \cite{Roth-07,AGN-long-07}\\
TA            & 2007-present & $860\times$yrs $  $&  &  & \cite{Fukushima-07} \\ \br
\end{tabular}
\end{center}
\end{table}

In case of ground arrays, the aperture is calculated in a
straight forward and model independent way, once the energy
threshold for CR detection and reconstruction is exceeded.  The
only uncertainty arises from the reconstruction of the landing
point of the shower, which again is safely reconstructed if only
showers within the geometry covered by the array are considered.
The situation is quite different for fluorescence telescopes.
Here, the maximum distance out to which showers can be observed
increases with increasing fluorescence light and thereby
increasing energy.  On the one hand, the growing aperture is
very attractive and cost effective, as it allows to observe more
showers at high energies.  On the other hand, the maximum
distance out to which EAS can be seen is directly related to the
signal-to-noise ratio in the light sensors of the cameras, the
varying atmospheric conditions, etc.  Moreover, an accurate
reconstruction of the CR energy requires the observation of the
position of the shower maximum, $X_{\rm max}$, in the
field-of-view of the telescopes.  This condition imposes a
sensitivity also to the primary mass and consequently also to the
hadronic interaction models employed in the aperture calculation.

\begin{figure}[b]
\begin{minipage}{18pc}
\includegraphics[width=15.5pc]{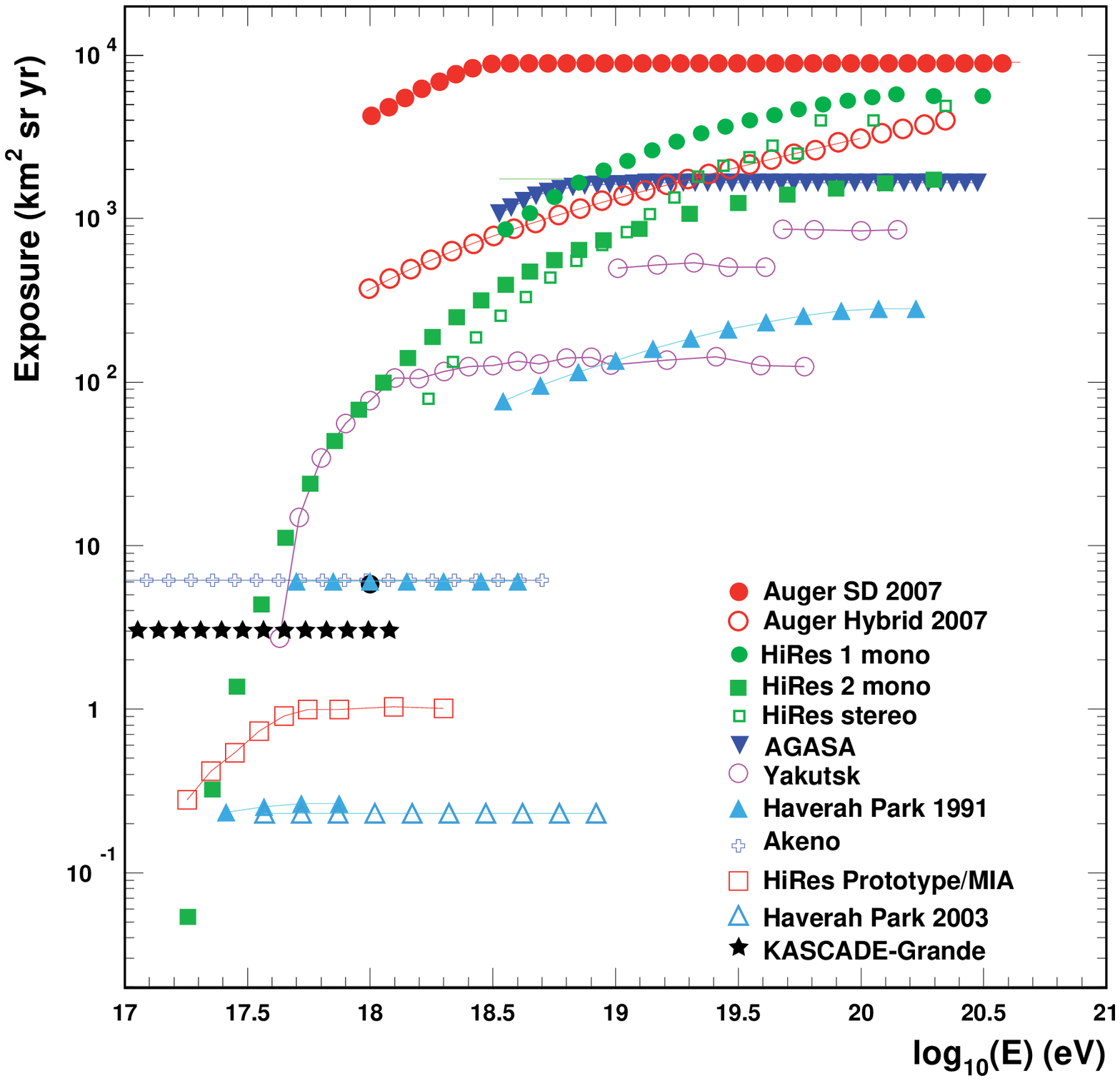}
\caption{\label{fig:exposures} Accumulated exposures of various
experiments.  The data and exposures are based on
Refs.~\cite{Bergman-07a,Hanlon-07,Bergman-Belz-07,Haungs-07,AGN-long-07}.}
\end{minipage}\hspace{2pc}%
\begin{minipage}{18pc}
\includegraphics[width=15.5pc]{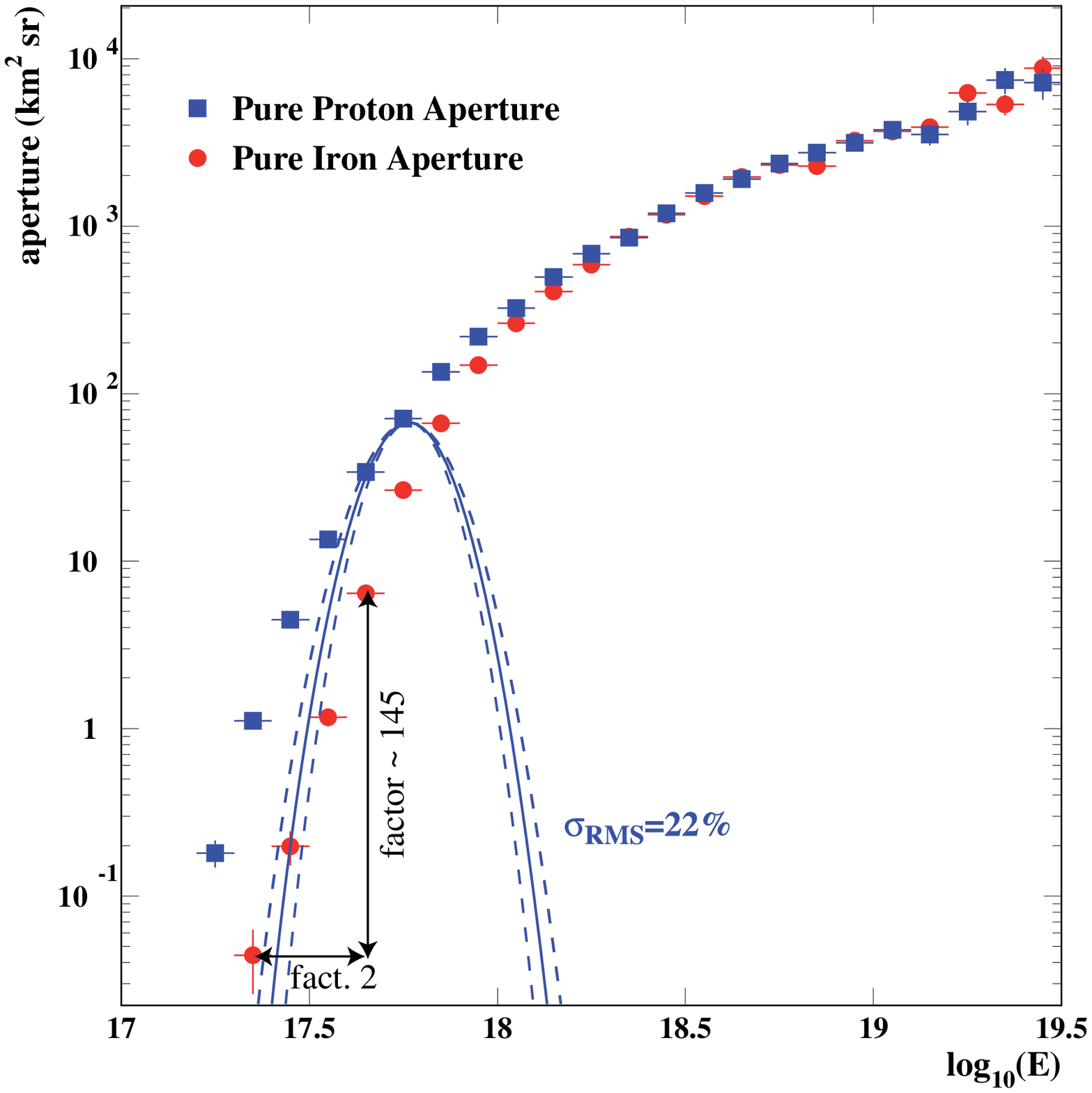}
\caption{\label{fig:exp-p-fe}HiRes apertures (from 
\cite{Abbasi-07b}) for
pure proton and iron simulations compared to the energy resolution.}
\end{minipage} 
\end{figure}

The effect of growing apertures in case of fluorescence
telescopes and constant apertures in case of ground arrays can be
seen in Fig.\,\ref{fig:exposures}.  Note also the difference in
the detection\-/reconstruction thresholds of HiRes-I and -II.
This is mostly because of the different ranges of elevation
angles viewed by HiRes-I ($3^{\circ}$-$17^{\circ}$) and -II
($3^{\circ}$-$31^{\circ}$) cameras: low energy showers, which can
- due to their low light level - only be observed near the
telescopes, reach their shower maximum above the field-of-view of
the cameras and thus cannot be reconstructed.
Fig.\,\ref{fig:exp-p-fe} compares in more details the aperture of
HiRes-II for p- and Fe-induced showers \cite{Abbasi-07b}.  Note
that the p- and Fe-apertures differ by more than a factor of 20
at $E\simeq 3\cdot 10^{17}$ eV! Thus, without knowledge of the
primary mass, the flux in this energy range is uncertain by a
factor of 20 (or more, dependent on the hadronic interaction
model used in the aperture calculation) and composition
measurements will be very biased.  We also illustrate the width
of the experimental energy resolution which indicates that the
rapid fall of the aperture distribution almost equals the energy
resolution function.  Fluctuations in the energy reconstruction
by only one standard deviation or a shift of the overall energy
scale within the uncertainty of the experiment causes changes
in the aperture (and thereby the CR-flux) by a factor of about 6!
Clearly, controlling all these uncertainties particularly at
energies below $10^{18}$~eV appears very difficult.

\section{The Energy Spectrum}

A very important step towards unveiling the origin of the sources
of UHECR is provided by measurements of the CR energy spectrum.
Four features are known to provide information about the CR
origin: the prominent {\em knee\/} at $E \simeq
4\cdot10^{15}$\,eV may signal the limiting energy of galactic CR
accelerators and/or the onset of diffusion losses from the
galaxy, the {\em second knee\/} at $E \simeq 10^{17}$\,eV, still
to be confirmed \cite{Kampert-07}, is considered to be caused by
the fading of the heavy galactic CR component, the {\em ankle\/}
at $E \simeq 4\cdot10^{18}$\,eV is either due to the onset of the
extragalactic CR component or due to energy losses of
extragalactic protons by $e^{+}e^{-}$ pair production in the CMB
\cite{Berezinsky-05}, and the {\em GZK cut-off\/} at $E\simeq
6\cdot10^{19}$\,eV \cite{GZK} is due to photo-pion production of
extragalactic protons in the CMB.

\begin{figure}[b]
\begin{minipage}{18.5pc}
\includegraphics[width=18.5pc]{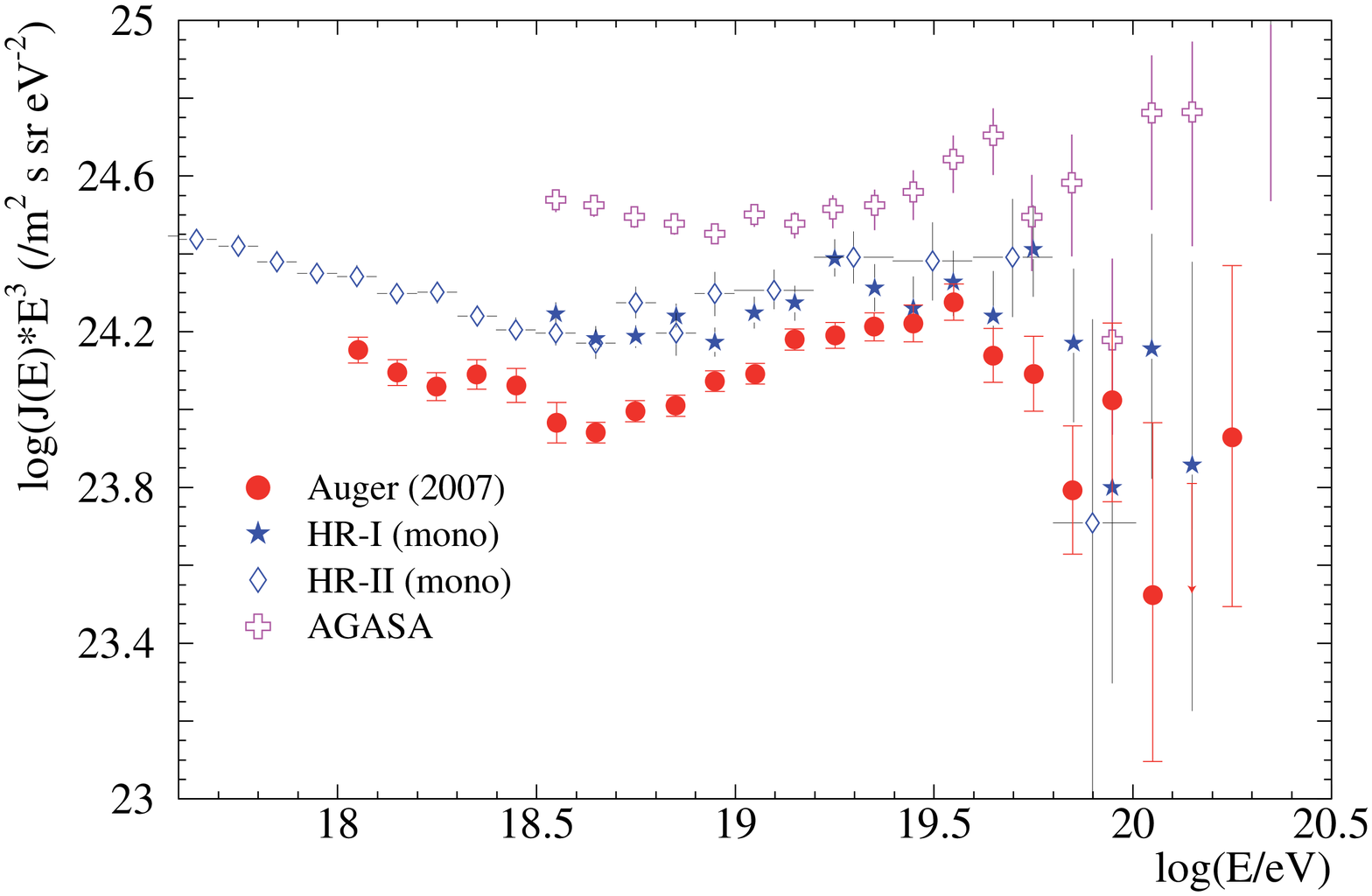}
\caption{\label{fig:e-spec-unscaled} Cosmic ray flux 
measurements (multiplied by $E^{3}$) from 
AGASA \cite{Takeda-03}, HiRes \cite{Bergman-07b}, and the PAO 
\cite{Roth-07}.}
\end{minipage}\hspace{1pc}%
\begin{minipage}{18.5pc}
\includegraphics[width=18.5pc]{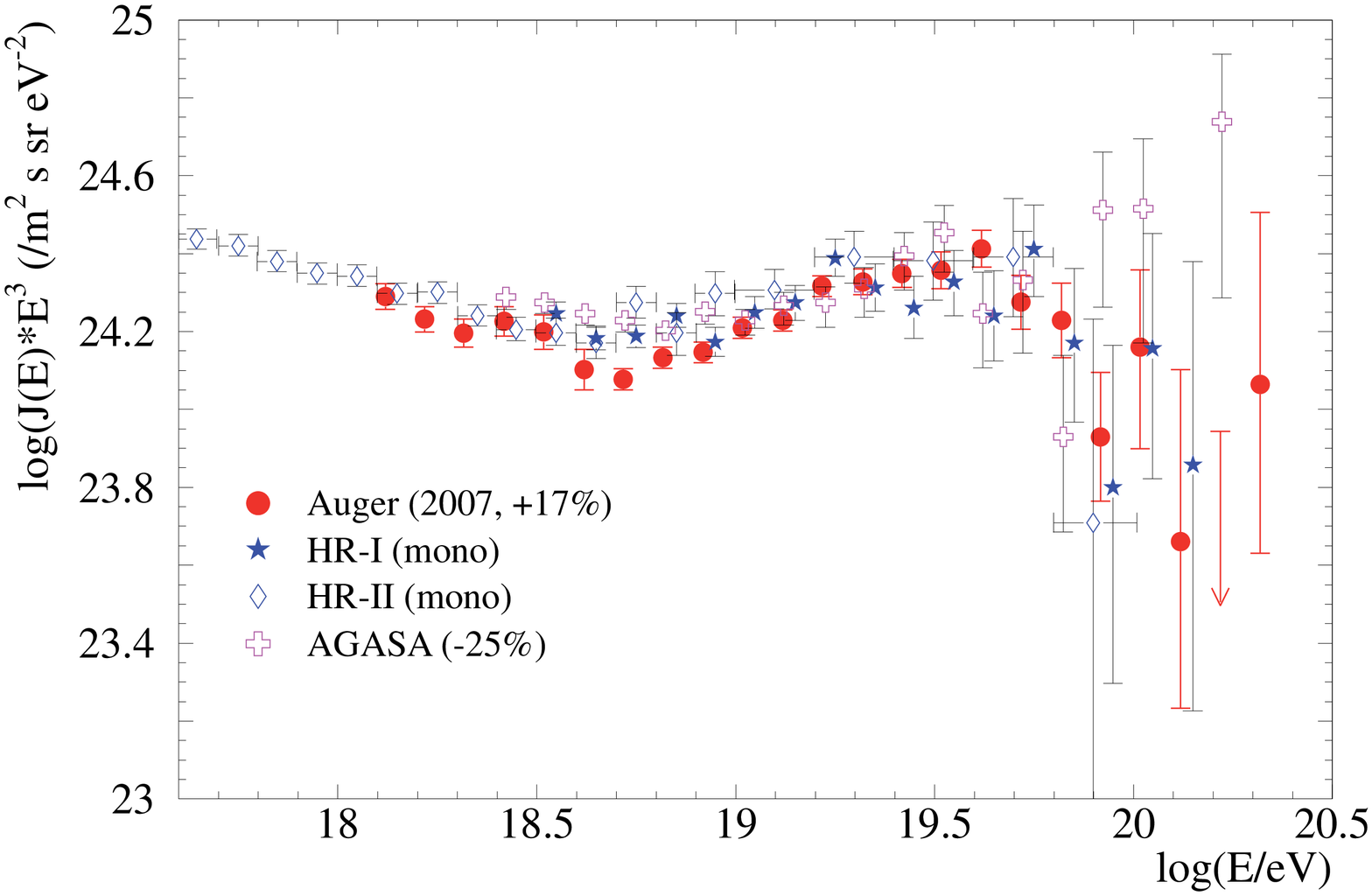}
\caption{\label{fig:e-spec-scaled}Same as
Fig.\,\ref{fig:e-spec-unscaled} but with the energy scale of
Auger and AGASA shifted by $+17$\,\% and $-25$\,\%,
respectively.}
\end{minipage} 
\end{figure}

Recent measurements of the CR energy spectrum by AGASA and HiRes
have yielded results which differ in their shape and overall
flux.  A comparison including data from the PAO as presented at
the ICRC 2007 is shown in Fig.\,\ref{fig:e-spec-unscaled}.
Generally, the error bars in such plots are of statistical nature
only and neglect systematic uncertainties in the determination of
the energy scale and exposure.  Typical uncertainties of the
energy scale are on the order of 20-25\,\%.  Ground arrays like
AGASA rely entirely on EAS simulations with their uncertainties
originating from the limiting knowledge of hadronic interactions
(total inelastic cross sections, particle multiplicities,
inelasticities, etc.).  CORSIKA simulations \cite{Drescher-04}
have shown that the muon density at ground predicted by different
hadronic interaction models differ by up to 30\,\%.  Fluorescence
telescopes, such as operated by HiRes and the PAO, observe the
(almost) full longitudinal shower development in the atmosphere.
In this way, the atmosphere is employed as a homogenous
calorimeter with an absorber thickness of 30 radiation lengths or
11 hadronic interaction lengths.  Corrections for (model
dependent) energy `leakage' into ground -mostly by muons and
neutrinos - are below 10\,\% and their uncertainties are only a
few percent.  As a consequence, fluorescence detectors provide an
energy measurement which is basically independent from hadronic
interaction models.  Uncertainties in the energy scale arise most
dominantly from the fluorescence yield in the atmosphere.
Several measurements have been made in the past, e.g.\ the Auger
Collaboration uses the fluorescence yield by Nagano et al.\
\cite{Nagano-04} and HiRes by Kakimoto et al.\
\cite{Kakimoto-96}.  This is an unpleasant situation, which by
itself may account for a $\sim 10$\,\% shift between the energy
scales of Auger and HiRes.  For this reason, major efforts have
been started to remeasure the fluorescence yield as a function of
temperature, pressure and humidity with high precision
\cite{fluorescence} in order to reduce this source of
uncertainty.

Taking benefit of the Auger hybrid detector, the Auger
Collaboration has used a clean set of hybrid data, in which EAS
have been detected simultaneously by at least one florescence eye
and the ground array, to calibrate their observatory
\cite{Roth-07}.  To evaluate the observed differences in the
energy scale of AGASA, HiRes and Auger in
Fig.\,\ref{fig:e-spec-unscaled}, an overall shift of $+17$\,\%
and $-25$\,\% has been applied to the Auger and AGASA energy
scale, respectively (c.f.\ Fig.\,\ref{fig:e-spec-scaled}).  Such
a shift remains well within the quoted uncertainties of the
experiments (particularly when accounting for the different
fluorescence yields used by HiRes and Auger) and yields a fairly
good agreement of the data points, except perhaps at $E \ga
10^{20}$\,eV for AGASA and in the ankle region which appears
sharper in the PAO data.

The GZK-like suppression is clearly visible in both the HiRes and
Auger data.  Using different statistical approaches, HiRes quotes
a significance of about 4.5 standard deviations and the PAO of
more than $6\sigma$.  For example, fitting a power law to the
Auger spectrum between $4\cdot10^{18}<E<4\cdot10^{19}$\,eV using
a binned likelihood method yields $\gamma=-2.69\pm0.02 {\rm
(stat)} \pm 0.06 {\rm (sys)}$.  Extrapolating this slope to
higher energies one expects $167 \pm 3$ and $35 \pm 1$ events at
$E > 4\cdot10^{19}$ and $10^{20}$\,eV, respectively, whereas 66
and 1 event are observed.  The observation of the GZK-effect 40
years after its prediction provides for the first time clear
evidence for an extragalactic origin of EHECRs.  Of course, this
interpretation is challenged if the sources would happen to run
out of acceleration power just at the value of the GZK threshold.
However, this would be a strange coincidence and in fact is not
supported by Pierre Auger data (see Sect.\ \ref{sec:corr}).

The shape of the energy spectrum around $10^{20}$\,eV carries
information about the distance distribution of CR sources and
their injection spectrum \cite{Yamamoto-07}.  However, more
statistics is required before firm conclusions can be drawn.
Answering the question about the origin of the ankle and
discriminating the $e^+e^-$ dip-model \cite{Berezinsky-05} from
the traditional ankle model cannot be done based on the energy
spectrum alone but requires measurements of the CR composition.

\section{Chemical Composition, Photon and Neutrino Limits}

\begin{figure}[t]
\begin{minipage}{18.5pc}
\includegraphics[width=18.5pc]{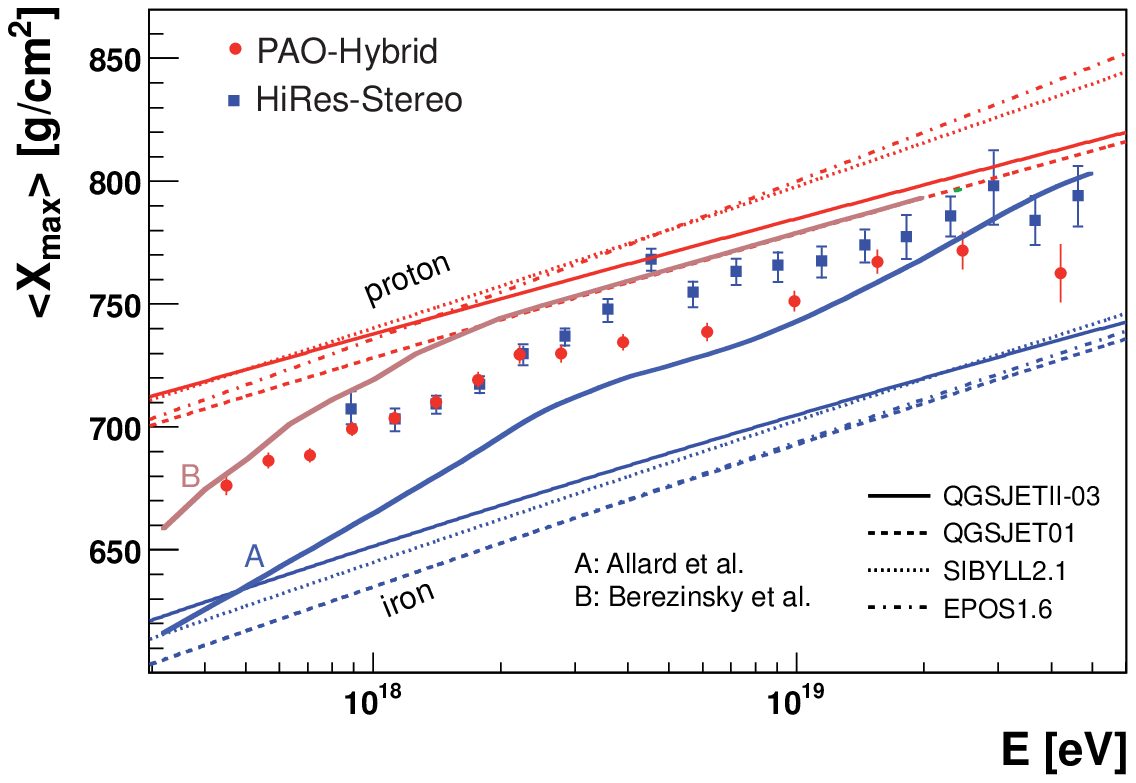}
\caption{\label{fig:Xmax} $\langle X_{\rm max} \rangle$ as a 
function of energy for HiRes Stereo \cite{Fedorova-07} and PAO 
hybrid data \cite{Unger-07} compared to proton and Iron 
predictions using different hadronic interaction models and 
different models of EHECR origin.}
\end{minipage}\hspace{1pc}%
\begin{minipage}{18.5pc}
\includegraphics[width=18.0pc]{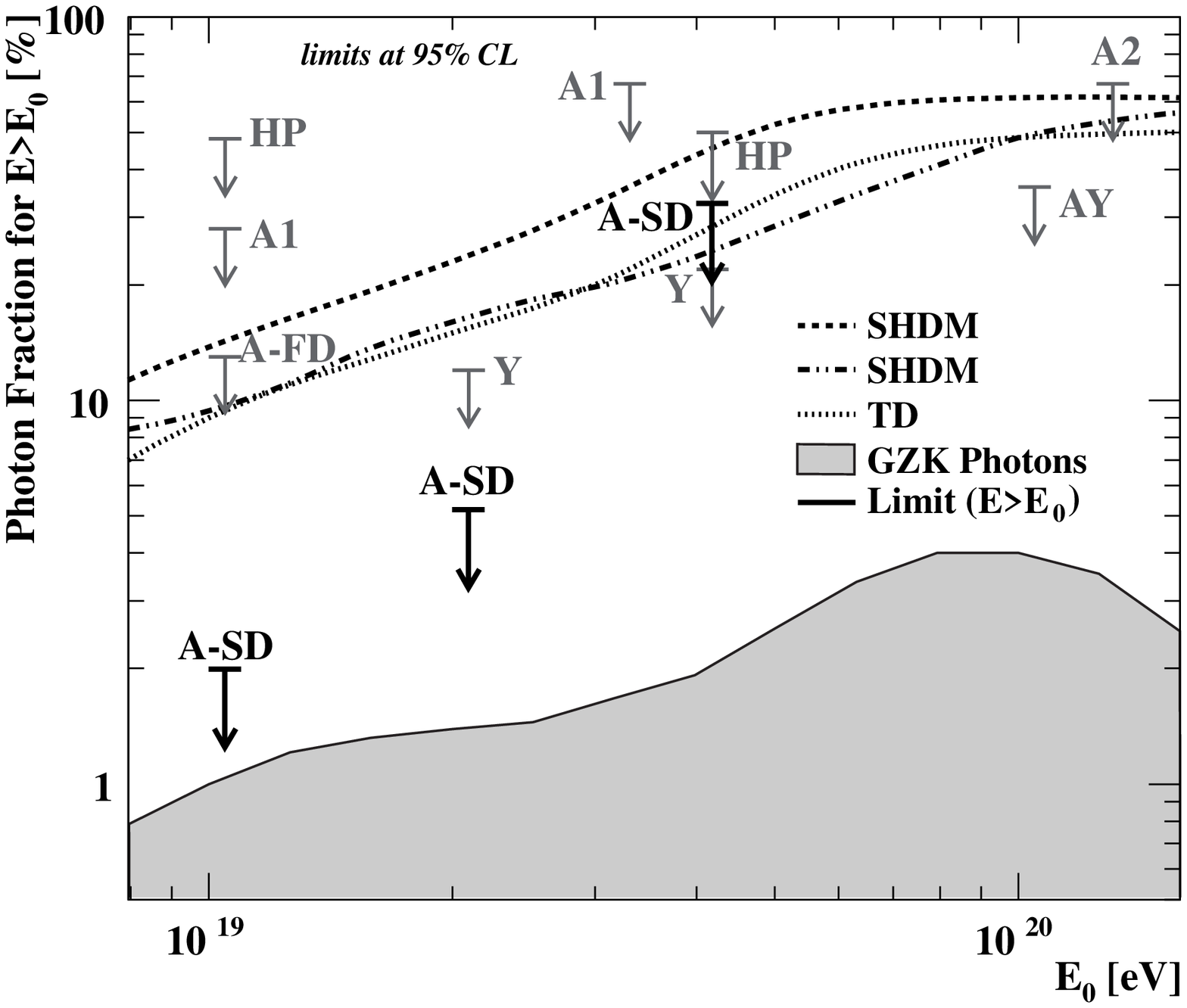}
\caption{\label{fig:photon-limits}Upper limits on the fraction 
of photons in the integral CR flux compared to predictions. For 
references see \cite{Auger-Photons-07}.}
\end{minipage} 
\end{figure}

As noted above, the mass composition of CRs allows to
discriminate models of UHECR origin and may be the only
measurement allowing to answer the question about the transition
from galactic to extragalactic CRs.  The Berezinsky dip-model
\cite{Berezinsky-05} predicts the transition taking place at
energies significantly below the ankle but requires a proton
dominant composition in the ankle region to make the
Bethe-Heitler pair production process work.  In the classical
picture, on the other hand, the ankle itself marks the transition
region and one expects a change from a heavy to light composition
at the ankle.  Unfortunately, the mass composition can be
inferred only indirectly from EAS experiments by making
assumptions about the hadronic interactions at the highest
energies.  In contrast to energy measurements, this model
dependence is true also for fluorescence detectors.  The key
observable here is the position of the shower maximum, $X_{\rm
max}$, which is directly observed by fluorescence telescopes and
can be inferred also from surface detector data.  New results
based on HiRes-Stereo and PAO hybrid data were reported at the
ICRC \cite{Unger-07,Fedorova-07}.  As can be seen from
Fig.\,\ref{fig:Xmax}, both data sets agree very well up to $\sim
3\cdot10^{18}$\,eV but differ slightly at higher energies.  The
differences between the two experiments is of the same order as
the differences observed between p- and Fe-predictions for
different hadronic interaction models.  With these caveat kept in
mind, both experiments observe an increasingly lighter
composition towards the ankle.  At higher energies, the HiRes
measurement yields a lighter composition than Auger.  Also shown
are predictions of $X_{\rm max}$ based on the QGSJET01 model for
the traditional G-EG transition \cite{Allard-07} (labelled ``A'')
and the Berezinsky dip-model \cite{Berezinsky-05} (labelled
``B'').  None of the two models appears to describe the
preliminary data well, but they demonstrate the power of such
measurements which will be particularly important in the energy
range $10^{17}$-$10^{18}$\,eV. The low energy upgrade HEAT with
additional high elevation telescopes, infill stations and muon
detectors in Auger \cite{HEAT}, the low energy extension TALE of
the TA \cite{TALE}, KASCADE-Grande \cite{Haungs-07,Bluemer-07},
and IceTop/IceCube \cite{Gaisser-07} will provide powerful data
in this energy range to provide a definite answer about the
G-EG-Transition already in the very near future.

Primary photons are easier to separate experimentally as they
penetrate deeper into the atmosphere, particularly at energies
above $10^{18}$\,eV. Their EAS development is also much less
affected by uncertainties of hadronic interaction models due to
the dominant electromagnetic shower component.  They are of
interest for several reasons: top-down models, invented to
explain the apparent absence of the GZK-effect in AGASA data,
predict a substantial photon flux at high energies
\cite{Risse-07}.  In the presence of a GZK effect, UHE photons
can act as tracers of the GZK process and provide relevant
information about the sources and propagation.  Moreover, they
can be used to obtain input to fundamental physics and UHE
photons could be used to perform EHE astronomy.

Experimentally, photon showers can be identified by their
longitudinal shower profile, most importantly by their deep
$X_{\rm max}$ position and low muon numbers.  Up to now, only
upper limits could be derived from various experiments, either
expressed in terms of the photon fraction or the photon flux.
Figure \ref{fig:photon-limits} presents a compilation of present
results on the photon fraction.  The most stringent limits are
provided by the Auger surface detector \cite{Auger-Photons-07}.
Current top-down models appear to be ruled out by the current
bounds.  This result can be considered an independent
confirmation of the GZK-effect seen in the energy spectrum.  It
will be very exciting to possibly even touch the flux levels
expected for GZK-photons ($p+\gamma_{CMB} \to p + \pi^{0} \to
p+\gamma\gamma$) after several years of data taking.

\begin{figure}[t]
\begin{minipage}{18.5pc}
\includegraphics[width=18.5pc]{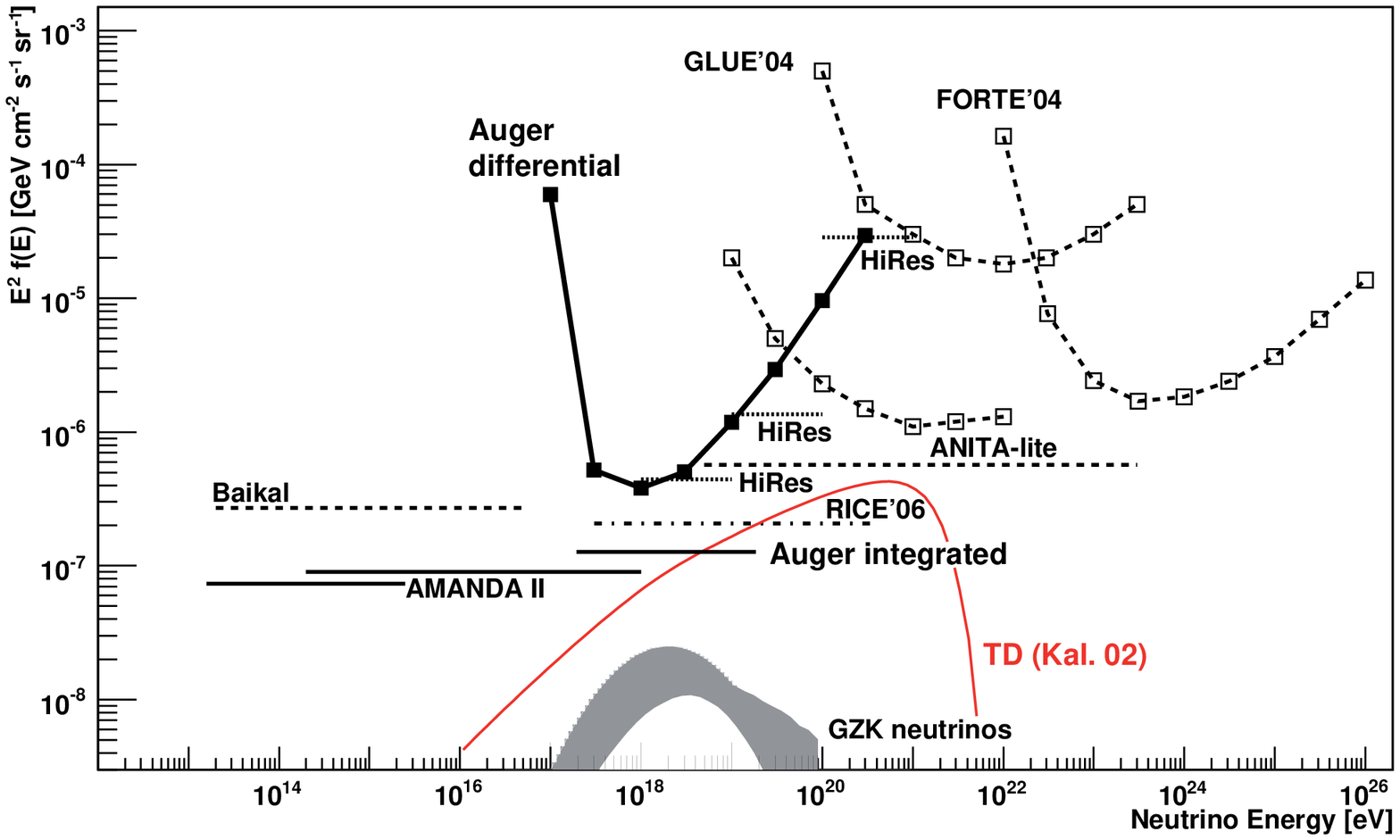}
\caption{\label{fig:neutrino-limits} Limits at the 90\,\% C.L. 
for a diffuse flux of $\nu_{\tau}$ assuming a 1:1:1 
ratio of the 3 neutrino flavors (\cite{Auger-Neutrino-07} and 
references therein) and predictions for a top-down model 
\cite{Kalashev-02}.}
\end{minipage}\hspace{1pc}%
\begin{minipage}{18.5pc}
\includegraphics[width=18.0pc]{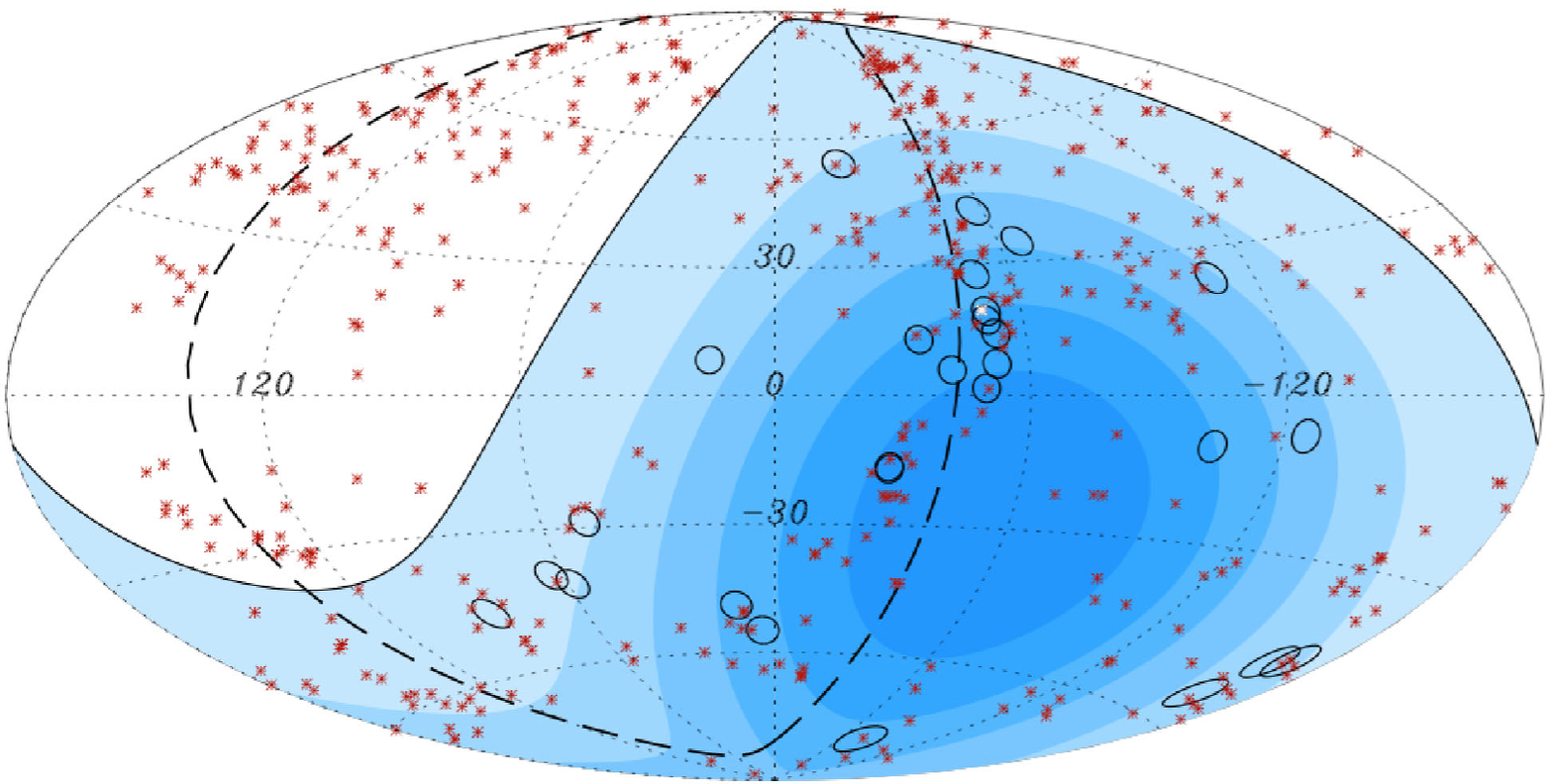}
\caption{\label{fig:AGN-sky}Aitoff projection of the celestial
sphere in galactic coordinates.  The positions of the AGN within
$D<71$\,Mpc (stars) and of the events with $E>57$\,EeV (circles)
are marked The colors indicate equal exposure
\cite{AGN-long-07}.}
\end{minipage} 
\end{figure}

The detection of UHE cosmic neutrinos is another long standing
experimental challenge.  All models of UHECR origin predict
neutrinos from the decay of pions and kaons produced in hadronic
interactions either at the sources or during propagation in
background fields.  Similarly to GZK-photons one also expects
GZK-neutrinos, generally called `cosmogenic neutrinos'.
Moreover, top-down models predict dominantly neutrinos at UHE
energies.  Even though neutrino flavors are produced at
different abundances, e.g.\ a 1:2 ratio of $\nu_{e}$:$\nu_{\mu}$
results from pion decay, neutrino oscillations during propagation
will lead to equal numbers of $\nu_{e}$, $\nu_{\mu}$, and
$\nu_{\tau}$ at Earth.  At energies above $10^{15}$\,eV,
neutrinos are absorbed within the Earth so that upgoing neutrino
induced showers cannot be detected anymore.  Only tau neutrinos
entering the Earth just below the horizon (Earth-skimming) can
undergo charged-current interactions to produce $\tau$ leptons
which then can travel several tens of kilometers in the Earth and
emerge into the atmosphere to eventually decay in flight
producing a nearly horizontal air shower above the detector.
Such showers can be searched for in ground arrays and
fluorescence detectors.  The absence of any candidates observed
in the detectors has been used to place upper limits on diffuse
neutrino fluxes.  As can be seen from
Fig.\,\ref{fig:neutrino-limits}, AMANDA and the PAO provide at
present the best upper limits up to energies of about
$10^{19}$\,eV and, similarly to the photons discussed above, they
already constrain top-down models and are expected to reach the 
level of cosmogenic neutrinos after several years of data taking.

\section{Arrival Directions and Correlations with AGN}
\label{sec:corr}

Recently, the Pierre Auger Collaboration reported the observation
of a correlation between the arrival directions of the highest
energy CRs and the positions of nearby AGN from the V\'eron-Cetty
catalogue at a confidence level of more than 99\,\%
\cite{Abraham-07,AGN-long-07}.  Since several claims about seeing
clustering of EHECRs were already made in the past with none of
them being confirmed by independent data sets, the Auger group
has performed an `exploratory' scan of parameters using an
initial data-set and applied these parameters to a new independent
data-set for confirmation.  With the parameters specified {\em a
priori\/} the analysis avoids the application of penalty factors
which otherwise would need to be applied for in {\em a
posteriori\/} searches.  The correlation has maximum significance
for CRs with energies greater than $5.7\cdot10^{19}$\,eV and AGN
at a distance less than $\sim 71$\,Mpc.  At this energy
threshold, 20 of the 27 events correlate within $3.2^{\circ}$
with positions of a nearby AGN.

Observing such kind of anisotropy can be considered the first
evidence for an extragalactic origin of the most energetic CRs
because none of any models of galactic origin even including a
very large halo would result in an anisotropy such as observed in
the data.  Besides this, the correlation parameters itself are
highly interesting as the energy threshold at which the
correlation becomes maximized matches the energy at which the
energy spectrum shows the GZK feature ($\sim 50$\,\% flux
suppression), i.e.\ CRs observed above this threshold need to
originate from within the GZK-horizon of $\sim 100$-$200$\,Mpc.
This number again matches (within a factor of two) the maximum
distance for which the correlation is observed!  Thus, the set of
the two parameters suggests that the suppression in the energy spectrum is indeed due to the GZK-effect, rather than to a limited energy of the accelerators. Thereby, the GZK-effect
acts as an effective filter to nearby sources and minimizes
effects from extragalactic magnetic field deflections. On top of this, it is also the large magnetic rigidity which helps to
open up the window for performing charged particle astronomy.

The correlation may tell us also about the strength of galactic
and extragalactic magnetic fields.  The galactic fields are
reasonably well known and one expects strong deflections for
particles arriving from nearby the galactic plane even at
energies of 60 EeV. And in fact, 5 of the 7 events that do not
correlate with positions of nearby AGN arrive with galactic
latitudes $|b|<12^{\circ}$.  The angular scale of the observed
correlation also implies that the intergalactic magnetic fields
do not deflect the CRs by more than a few degrees and one can
constrain models of turbulent magnetic fields to $B_{\rm
rms}\sqrt{L_{c}} \le 10^{-9}{\rm ~G}\sqrt{\rm Mpc}$ within the
GZK horizon assuming protons as primary particles
\cite{AGN-long-07}.

\section{Concluding Remarks}

Remarkable progress has been made in cosmic ray physics at the
highest energies, particularly by the start-up of the (still
incomplete) Pierre Auger Observatory.  The event statistics
above $10^{19}$\,eV available by now allows detailed comparisons
between experiments and indicates relative shifts of their energy
scales by $\pm 25$\,\%.  Given the experimental and theoretical
difficulties in measuring and simulating extensive air showers at
these extreme energies, this may be considered a great success.
On the other hand, knowing about overall mismatches of the energy
scales between experiments may tell us something.  Clearly, in
case of fluorescence detectors better measurements of the
spectral and absolute fluorescence yields and their dependence on
atmospheric parameters are needed and will hopefully become
available in the very near future \cite{fluorescence}.  This
should furnish all fluorescence experiments with a common set
of data.  Differences in the calibration between surface
detectors and fluorescence telescopes, best probed by hybrid
experiments like Auger and TA, may then be used to test the
modelling of EAS. The muon component at ground, known to be very
sensitive to hadronic interactions at high energies
\cite{Drescher-04}, could in this way serve to improve hadronic
interaction models in an energy range not accessible at man-made
accelerators.  In fact, several studies (e.g.\ \cite{Engel-07})
indicate a deficit of muons by 30\,\% or more in interaction
models like QGSJET.

The energy scale is of great importance also for the AGN
correlation discussed in the previous section. As shown in
\cite{AGN-long-07}, the correlation sets in abruptly at an
(Auger) threshold energy of about 57 EeV. Already a downshift in energy
by 17\,\% (the mismatch between Auger and HiRes) would weaken the
signal by more than 3 orders of magnitude to make it basically
disappear.  Thus, verification of the correlation signal by HiRes
or AGASA would need to be done for a threshold energy (on their
scale) of 67 EeV and 85 EeV, respectively.  In this energy range,
HiRes observes a spectral slope of $\gamma=5.1\pm0.7$
\cite{Bergman-07b}, i.e.\ the number of events available for a
correlation analysis would, according to Table \ref{tab:expts},
drop to about 12 (HR-I) and 4 (HR-II) when taking the rise of the
apertures into account.  This would amount to about half the
statistics of Auger, well in agreement with the quoted exposures.
Unfortunately, the angular resolution of monocular reconstruction
is by far too poor for such a test.  Only stereo data could
provide the required angular resolution. However, in this case
the expected statistics of about 5 events above threshold (based
upon the numbers and exposures given above) appears too small for
any verification.

In fact, the distance parameter of the correlation of 71 Mpc may
indicate a mismatch of the energy scale: For protons above 57 EeV
the GZK horizon would be 200 Mpc \cite{Harari-06} but already for
20\,\% higher energy it would shrink by more than a factor of two
to become consistent to the correlation parameter.  Another
puzzling feature is the observed small deflection of particles
which suggests dominantly protons as primaries.  Note that 90\,\%
of the events (20/22) off the galactic plane are correlated to
within $\sim 3^\circ$ which AGN positions which is very unlikely
for heavy nuclei.  On the other hand, the elongation curves in
Fig.\,\ref{fig:Xmax} suggests an admixture of heavy nuclei by
more than 10\,\%.  This may be related again to imperfections of
the hadronic interaction models used for comparison in
Fig.\,\ref{fig:Xmax}.

All of this tells us that the near future will be highly
exciting: The question of the energy scales will soon be settled
and more detailed comparisons between experiments will become
possible. The shape of the energy spectrum in the GZK region will
tell us about the source evolution, the composition in the ankle
region will answer the question about the G-EG transition,
observations of cosmogenic photons and neutrinos are in reach and
in case of neutrinos will probe the GZK effect over larger
volumes, the correlations will be done with better statistics,
with improved search techniques and with more appropriate source
catalogues and source selection parameters to tell us about
source densities, and the true sources of EHECRs.  Very important
to note is that different pieces of information start to mesh and
are being accessed from different observational techniques and can
be cross-checked: {\em The big picture is being painted!}

Given the scientific importance of this, it would be a mistake to
have only one observatory - even when operated as a hybrid
detector - taking data.  The TA project and its extensions will
be very important particularly in the sub-GZK range but,
unfortunately, will be too small to collect sufficient statistics
at the highest energies.  Auger-North will be imperative here and
needs immediate vigorous support.  The next generation experiment
JEM EUSO to be mounted at the Exposed Facility of Japanese
Experiment Module JEM EF will potentially reach much larger
exposures but still faces many experimental challenges to be
addressed.

\vspace*{-5mm}
\subsection*{Acknowledgement}
I would like to thank the organizers of the excellent TAUP
meeting in Sendai for the invitation to give a review talk at
this exciting time. Also, its a pleasure to thank many of my colleagues for stimulating 
discussions. The German Ministry for Research and Education 
(BMBF) and the Deutsche Forschungsgemeinschaft (DFG) are 
gratefully acknowledged for financial support.

\end{document}